\documentclass[aps,pre,reprint,groupedaddress, showpacs]{revtex4-1}

\usepackage[version=3]{mhchem}
\usepackage{balance}
\usepackage{graphicx}
\usepackage{lastpage}
\usepackage{mathptmx}
\usepackage{amssymb}
\usepackage{upgreek}
\usepackage{dsfont}
\usepackage{epstopdf}
\usepackage{subfigure}
\usepackage[T1]{fontenc}
\usepackage{bm}
\usepackage{tipa}

\begin{document}


\title{Effect of body deformability on microswimming}



\author{Jayant Pande\textsuperscript{1}, Laura Merchant\textsuperscript{1,2}, Timm Kr{\" u}ger\textsuperscript{3}, Jens Harting\textsuperscript{4,5} and Ana-Sun{\v c}ana Smith\textsuperscript{1,6,*}}


\affiliation{\textsuperscript{1}PULS Group, Department of Physics and Cluster of Excellence:~EAM, Friedrich-Alexander University Erlangen-Nuremberg, Erlangen, Germany; \textsuperscript{2}School of Physics and Astronomy, University of St. Andrews, St. Andrews, Scotland; \textsuperscript{3}School of Engineering, The University of Edinburgh, United Kingdom; \textsuperscript{4}Department of Applied Physics, Eindhoven University of Technology, Eindhoven, The Netherlands; \textsuperscript{5}Helmholtz Institute Erlangen-Nuremberg for Renewable Energy (IEK-11), Nuremberg, Germany; \textsuperscript{6}Division of Physical Chemistry, Ru\dj{}er Bo\v skovi\'c Institute, Zagreb, Croatia.}





\email[]{smith@physik.uni-erlangen.de}

\date{\today}

\begin{abstract}
In this work we consider the following question: given a mechanical microswimming mechanism, does increased deformability of the swimmer body hinder or promote the motility of the swimmer? To answer this we study a microswimmer model composed of deformable beads connected with springs. We determine the velocity of the swimmer analytically, starting from the forces driving the motion and assuming that the oscillations in the effective radii of the beads are known and are much smaller than the radii themselves. We find that to the lowest order, only the driving frequency mode of the surface oscillations contributes to the swimming velocity, and that this velocity may both rise and fall with the deformability of the beads depending on the spring constant. To test these results, we run immersed boundary lattice Boltzmann simulations of the swimmer, and show that they reproduce both the velocity-promoting and velocity-hindering effects of bead deformability correctly in the predicted parameter ranges. Our results mean that for a general swimmer, its elasticity determines whether passive deformations of the swimmer body, induced by the fluid flow, aid or oppose the motion.
\end{abstract}


\maketitle


\section{Introduction}

The study of microswimmer motion has gained a lot of impetus recently, driven in equal measure by advances in experimental technology,~\cite{Dreyfus:2005:Nature, Ishiyama:2001:SAA-P, Benkoski:2011:JMaterChem, Li:2012:JApplPhys, Breidenich:2012:SoftM, Baraban:2012:ACSNano, Baraban:2012:SoftM, Keim:2012:PhysFluids, Theurkauff:2012:PRL, Sanchez:2011:Science, Liao:2007:PhysFluids, Leoni:2009:SoftM} numerical methods,~\cite{Pooley:2007:PRL, Elgeti:2013:PNAS, Zoettl:2012:PRL, Pickl:2012:JoCS, Swan:2011:PhysFluids} and theoretical modelling.~\cite{Guenther:2008:EPL, Becker:2003:JFM, Ledesma-Aguilar:2012:EurPhysJE, Avron:2005:NewJPhys, Friedrich:2012:PRL, Fuerthauer:2013:PRL, Leoni:2010:PRL} The increased attention has served to highlight the dazzling variety of ways in which nature accomplishes the difficult task of achieving non-reversibility of motion, as is required for propagation at low Reynolds numbers,~\cite{Purcell:1977:AmJPhys} despite relative sparseness of degrees of motile freedom. 

In spite of the great diversity, a great number of swimmers can be classified as mechanical, as their motion is driven by different parts of their body moving in coordinated yet asymmetric ways, leading to a corresponding asymmetry in the fluid flow surrounding them and thereby motion. In nature, this class of microswimmers is predominant.~\cite{Chapman-Andresen:1976:CRC, DiLuzio:2005:Nature, Kreutz:2012:JEukaryotMicrobiol, Fisher:2014:ProcRSocB, Ueki:2016:PNAS} For artificial swimmers, chemically driven mechanisms are as popular as mechanical ones.~\cite{Golestanian:2005:PRL, Howse:2007:PRL, Cordova-Figueroa:2008:PRL, Erb:2009:AdvMat, Ebbens:2010:SoftM, Buttinoni:2012:JPhysCondMat, Ghosh:2013:PRL}

An important consideration in mechanical microswimming is the way elastic forces interact with the fluid and any external forces present in determining the motion. In the literature different aspects of these forces that have been studied are the elasticity of the swimming mechanism (such as those involving flagella and cilia),~\cite{Elgeti:2013:PNAS, Downton:2009:EPL} the influence of flexible surrounding walls if the motion occurs close to them,~\cite{Ledesma-Aguilar:2013:PRL, Dias:2013:PhysFluids} and even the elasticity (or the viscoelasticity) of the ambient fluid.~\cite{Chaudhury:1979:JFM,Fu:2010:EPL, Teran:2010:PRL, Keim:2012:PhysFluids, Spagnolie:2013:PRL, Gagnon:2013:EPL, Thomases:2014:PRL, Riley:2015:JTB, Li:2015:JFM, Wrobel:2016:JFM} In addition, the advantage of having elastic bodies has been investigated for the forced motion of micro-bodies, such as capsules driven through constrictions.~\cite{Kusters:2014:PRE}

There is, however, one facet of elasticity and its role in microswimming that has so far not been adequately considered. This has to do with the fluid pressure that swimmers with elastic, yielding bodies face and the deformations of shape that they undergo in response. These deformations, which we term \textit{passive} in contrast to \textit{active} shape deformations which occur under the swimmer's own agency and drive its motion,~\cite{Avron:2005:NewJPhys, Farutin:2013:PRL, Wu:2015:PRE} modify continuously both the friction coefficient of the swimmer and the fluid flow around it, and thereby influence the swimming. That small changes in morphology can dramatically alter swimming behaviour is illustrated by the fact that for some bacteria, size changes of $0.1$ $\mu$m can cause a $100000$ times higher energetic cost of motion.~\cite{Mitchell:2002:AmNat} 

The question of how passive deformations affect a microswimmer's motility is our focus of study in this paper. In nature this issue could be significant for a class of micro-organisms such as the euglenid \textit{Eutreptiella gymnastica}~\cite{Throndsen:1969:NJB} which undergo a process called \textit{metaboly}, wherein their bodies deform constantly during the swimming motion which is driven mainly by flagella or similar swimming appendages.~\cite{} In the literature both the mechanism and the function of metaboly--whether it is beneficial for locomotion, food capture, or some other purpose--remain unclear, although there is evidence to suggest that metaboly can be an efficient mode of motility in simple fluids and may become important in granular or complex media.~\cite{Arroyo:2012:PNAS} Our results here support this thesis, by showing that it is possible for passive shape deformations to enhance the swimming motility.

Apart from playing a role in biology, surface deformability of microswimmers can potentially increase the efficiency of artificial microswimming machines. In the literature many mechanical microswimmer models exist, both theoretical and experimental, which make use of periodically-beating elastic components connected to rigid bodies.~\cite{Purcell:1977:AmJPhys, Najafi:2004:PRE, } Our study answers the question of whether similar but better models can be designed by replacing the rigid bodies by deformable ones--more precisely, when and under what conditions surface deformability enhances these swimmers' motility.

To explore these questions we choose as our model of study a swimmer composed of deformable beads connected in a line through springs (explained in detail in the next section), an extension of the three-sphere swimmer introduced by Najafi and Golestanian.~\cite{Najafi:2004:PRE} The latter swimmer is a popular one in the field owing to its simplicity,~\cite{} and as we show in this paper, our extension of it retains this simplicity, thereby allowing us to study it analytically and through simulations, while imparting to it the features that are needed to investigate the influence of passive deformations on microswimming. In our theoretical model we assume that the surface deformations are small, and in this limit we find that both an increase and decrease in the swimming velocity are possible when the deformability of the beads increases. We christen these responses of the velocity as defining the `deformability-enhanced' and `deformability-hindered' regimes of swimming, respectively. For identical oscillation amplitudes of the surfaces of the three beads, the regime in which the swimming occurs is decided by the spring constant, independent of the bead deformability itself. 

In addition to the theory, we examine our swimmer model through fully-resolved immersed boundary lattice-Boltzmann simulations, where the stroke of the swimmer as well as the deformations of the beads occur in response to the different forces in the system. The simulation results confirm both the deformability-enhanced and deformability-hindered regimes from theory for all the values of the parameters investigated.

\section{Theoretical model}

\begin{figure}
\centering
\includegraphics[width=0.48\textwidth]{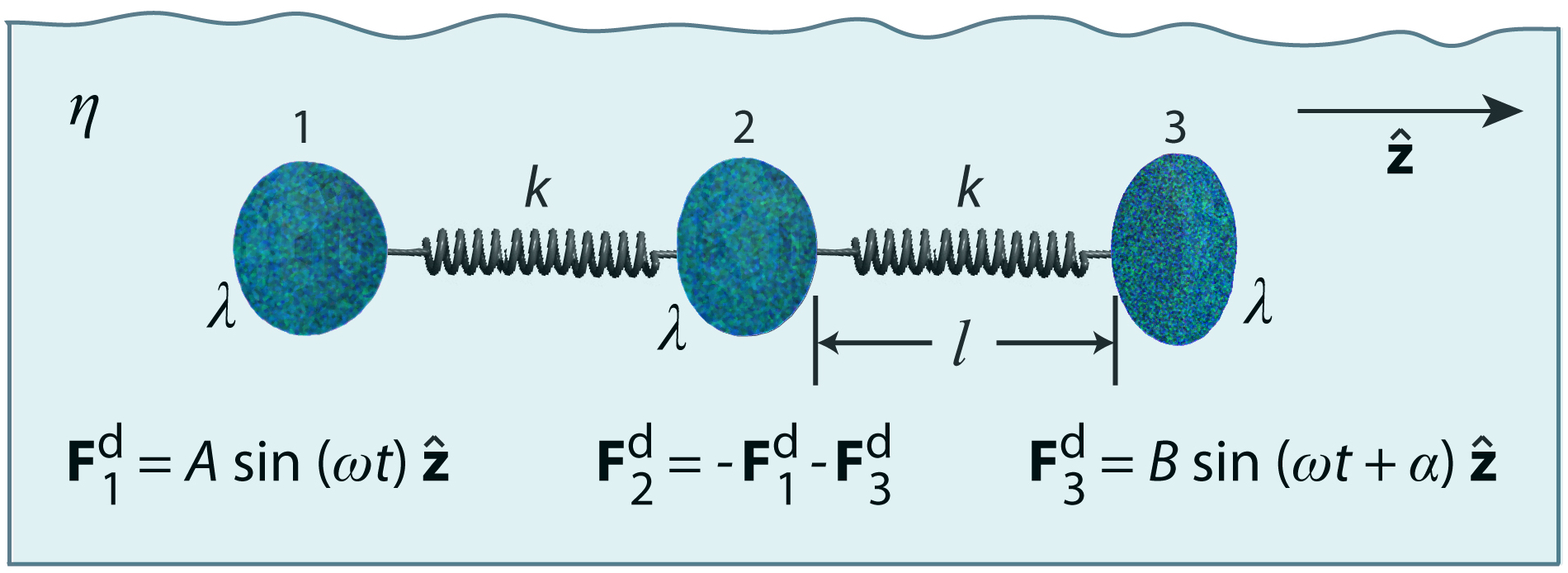}
\caption{(Colour online) A swimmer composed of three deformable vesicles connected by springs. The driving force on each vesicle is stated underneath it.}\label{fig:swimmer}
\end{figure}

The microswimmer is made up of three deformable beads, or membranes, aligned collinearly along the $z$-axis with harmonic springs connecting them (Fig.~\ref{fig:swimmer}). For simplicity, we take the springs to have equal stiffness constants $k$ and equal rest lengths $l$ (Fig.~\ref{fig:swimmer}), with $l$ taken to be much larger than the bead dimensions. The beads are driven by known forces which sum to zero at all times, in order to model autonomous swimming. 
The forces driving the motion act on the centres of the beads at all instants and are assumed to be known and of the form
\begin{align}\label{eq:dr_forces}
&\mathbf F_1^\text{d}(t) = A \sin\left(\omega t\right) \mathbf{\hat{z}};\ \ \ \ \mathbf F_2^\text{d}(t) = - \mathbf F_1^\text{d}(t) - \mathbf F_3^\text{d}(t);\nonumber\\
&\mathbf F_3^\text{d}(t) = B \sin\left(\omega t + \alpha\right) \mathbf{\hat{z}},\text{ with }\alpha \in[-\pi, \pi].
\end{align}
$A$ and $B$ specify the amplitudes of the driving forces $\mathbf F_1^\text{d}(t)$ and $\mathbf F_3^\text{d}(t)$ which are applied to the leftmost and the rightmost beads, respectively, along the $\mathbf{\hat{z}}$-direction. The driving frequency is $\omega$, and $\alpha$ is the phase difference between $\mathbf F_1^\text{d}(t)$ and $\mathbf F_3^\text{d}(t)$. 

In previous investigations of microswimming,~\cite{Pande:2015:SoftM, Pande:2016:PNAS} we have employed a similar swimmer model but with rigid beads. In order to incorporate the deformability of the beads into the model, we here assume that the change in shape of each deformable bead during a swimming cycle gives rise to a corresponding time-varying effective friction coefficient, which is \textit{a priori} known. This assumed knowledge of the effective bead shapes is crucial for an analytical solution of the problem, since the complicated two-way interactions between the fluid flow and the membrane deformations are effectively decoupled. A different possible approach would be to calculate the deformations from first principles, but this is a very complicated problem, requiring very likely a numerical solution. We follow such a numerical approach in our lattice Boltzmann simulations of the swimmer, discussed in the next section.

The effective friction coefficient $\lambda_i(t)$ for the $i^\text{th}$ bead is given by
\begin{equation}
\lambda_i(t) = \dfrac{\gamma_i(t)}{6\pi\eta}.
\end{equation}
Here $\gamma_i(t)$ is the Stokes drag coefficient of the $i^\text{th}$ bead, clearly dependent upon the instantaneous shape of the bead, and $\eta$ is the dynamic viscosity of the fluid, assumed to be incompressible and Newtonian.

The motion occurs in the Stokesian regime, so that the fluid flow is governed by the equations
\begin{align}
\eta\nabla^2\mathbf u\left(\mathbf r, t\right) - \nabla p\left(\mathbf r, t\right) + \mathbf f\left(\mathbf r, t\right)&=0,\text{ and}\\
\nabla\cdot\mathbf u&=0,
\end{align}
where $\mathbf u\left(\mathbf r, t\right)$ and $p\left(\mathbf r, t\right)$ are the velocity and the pressure, respectively, of the fluid at the point $\mathbf r$ at time $t$. The force density $\mathbf f\left(\mathbf r, t\right)$ acting on the fluid is given by
\begin{equation}
\mathbf f\left(\mathbf r, t\right)=\sum\limits_{i=1}^3\left(\mathbf F^\text{d}_i(t) + \mathbf F^\text{s}_i(t)\right)\delta\left(\mathbf r - \mathbf R_i(t)\right),
\end{equation}
where the index $i=1, 2, 3$ denotes the $i$-th bead placed at the position $\mathbf R_i(t)$ subject to a driving force $\mathbf F^\text{d}_i(t)$ and a net spring force $\mathbf F^\text{s}_i(t)$. The latter can be written as
\begin{align}\label{eq:F^s}
& \mathbf F^\text{s}_i(t)  = \sum\limits_{j \neq i}^3 \mathbf G\left(\mathbf R_i(t) - \mathbf R_j(t)\right), \text{ with} \nonumber \\
& \mathbf G\left(\mathbf R_i(t) - \mathbf R_j(t)\right)  = -k \left(\dfrac{\mathbf R_i(t) - \mathbf R_j(t)}{|\mathbf R_i(t) - \mathbf R_j(t)|}\right)\left(|\mathbf R_i(t) - \mathbf R_j(t)| - l\right)
\end{align}
if $i$ and $j$ denote neighbouring beads, and $\mathbf G\left(\mathbf R_i(t) - \mathbf R_j(t)\right) = \mathbf 0$ otherwise. 
Assuming no slip at the fluid-bead interfaces, the instantaneous velocity $\mathbf v_i(t)$ of each bead~\cite{Doi:1988:OUP} is given by
\begin{align}\label{eq:v_i}
\mathbf v_i(t) = \dot{\mathbf R}_i(t) = & \left(\mathbf F^\text{d}_i(t) + \mathbf F^\text{s}_i(t)\right)/\left(6 \pi \eta \lambda_i(t)\right) \nonumber\\
					& + \sum\limits_{j \neq i}^3\mathbf T\left(\mathbf R_i(t) - \mathbf R_j(t)\right)\cdot\left(\mathbf F^\text{d}_i(t) + \mathbf F^\text{s}_i(t)\right),
\end{align}
where $\mathbf T\left(\mathbf r\right)$ is the Oseen tensor.~\cite{Happel:1965:P-H, Oseen:1927:LeipzigAV} Here in $\dot{\mathbf R}_i(t)$ and throughout the paper, a dot over a variable denotes its derivative with respect to time.

As in the rigid body case,~\cite{Pande:2016:PNAS} and as borne out by simulations, the bead positions in the steady state take the form
\begin{equation}\label{eq:R_i}
\mathbf R_i(t)=\mathbf S_{i0} +  \boldsymbol\xi_i(t) + \mathbf v_\text{def}\,t,
\end{equation}
where $\boldsymbol\xi_i(t)$ denotes small sinusoidal oscillations, which will be taken as perturbation variables around the equilibrium configuration $\mathbf S_{i0}$ of the swimmer. The swimmer moves with a cycle-averaged velocity $\mathbf v_\text{def}$, where the subscript denotes the `deformable' bead case.

As stated earlier, we take $\lambda_i(t)$ to be known functions. Given any periodic form for the same, they can be expressed in Fourier series in $\omega t$ of the form
\begin{equation}\label{eq:lambda_i}
\lambda_i(t) = a_i + \sum\limits_{n = 1}^{\infty}b_i^n \sin\left(n\omega t + \phi_i^n\right),\ \ i = 1, 2, 3,
\end{equation}
where $a_i$ is the effective friction coefficient of the mean shape of the $i^\text{th}$ bead, and $b_i^n$ and $\phi_i^n$ are the amplitude and the phase shift of the contribution from the $n^\text{th}$ frequency mode. The condition of weak deformability is satisfied by requiring $b_i^n/a_j \ll 1$ and $b_i^n/|\boldsymbol\xi_j(t)| \ll 1$ to hold for all $i$, $j$, and $n$ and time $t$. 

Under this assumption, we can write the mean swimming velocity $\mathbf v_\text{def}$ as the average velocity of the centre of hydrodynamic reaction $\mathbf C(t)$ of the swimmer (which takes the place of the centre of mass for low Reynolds number flows)~\cite{Happel:1965:P-H} over one swimming cycle, given by
\begin{equation}\label{eq:v_def}
\mathbf v_\text{def} = \dfrac{\omega}{2\pi}\int\limits_0^{2\pi/\omega}\mathrm dt\dot{\mathbf{C}}(t) = \dfrac{\omega}{2\pi}\int\limits_0^{2\pi/\omega}\mathrm dt\sum\limits_{i = 1}^{3}\lambda_i(t)\dot{\mathbf{R}}_i(t)/\sum\limits_{i = 1}^{3}\lambda_i(t).
\end{equation}


Expanding $\dot{\mathbf{C}}(t)$ from Eq.~(\ref{eq:v_def}),
\begin{align}\label{eq:Cdot}
\dot{\mathbf{C}}(t) = &\dfrac{\sum\limits_{i = 0} \left(a_i + \sum\limits_{n = 1}^{\infty}b_i^n \sin\left(n\omega t + \phi_i^n\right) \dot{\mathbf{R}}_i(t)\right)}{\left(\sum\limits_{j = 1}^3 a_j\right)\left(1 + \sum\limits_{k = 1}^3\sum\limits_{m = 1}^\infty \bar{b}_k^m \sin(m\omega t + \phi_k^m)\right)},
\end{align}
where $\bar{b}_k^m = b_k^m/\sum\limits_{l = 1}^3 a_l$. Here $\dot{\mathbf R_i}(t)$ can be written, using Eqs.~(\ref{eq:F^s})-(\ref{eq:lambda_i}), as
\begin{align}\label{eq:Rdot}
&\dot{\mathbf R_i}(t) = \dfrac{\mathbf F_i^\text{d}(t) + \sum\limits_{j \neq i}^3 \mathbf G\left(\mathbf R_i(t) - \mathbf R_j(t)\right)}{6 \pi \eta \left(a_i + \sum\limits_{n = 1}^{\infty}b_i^n \sin\left(n\omega t + \phi_i^n\right)\right)}\nonumber\\
 & + \sum\limits_{j \neq i}^3 \mathbf T\left(\mathbf R_i (t) - \mathbf R_j (t)\right)\cdot\left(\mathbf F_j^\text{d}(t) + \sum\limits_{k \neq j}^3 \mathbf G\left(\mathbf R_i(t) - \mathbf R_j(t)\right)\right).
\end{align}

Following the method of Felderhof,~\cite{Felderhof:2006:PhysFluids} we expand the functions of $\left(\mathbf R_i(t) - \mathbf R_j(t)\right)$ above in terms of series of the variables $\left(\boldsymbol\xi_i(t) - \boldsymbol\xi_j(t)\right)$ centred around the equilibrium configuration, which can be taken to be the configuration at time $t = 0$. The different variables expanded to the first order in $\left(\boldsymbol\xi_i(t) - \boldsymbol\xi_j(t)\right)$ are
\begin{align}\label{eq:G_ij}
&\mathbf G\left(\mathbf R_i(t) - \mathbf R_j(t)\right) = \nonumber \\
&\mathbf G\left(\mathbf R_i(0) - \mathbf R_j(0)\right)\nonumber + \left.\dfrac{\partial \mathbf G\left(\mathbf R_i(t) - \mathbf R_j(t)\right)}{\partial \mathbf R_i(t)}\right|_{t = 0}  \cdot \left(\boldsymbol\xi_i(t) - \boldsymbol\xi_j(t)\right) \nonumber \\
& = \mathbf H_{ij} \cdot \left(\boldsymbol\xi_i(t) - \boldsymbol\xi_j(t)\right), \text{ with } \mathbf H_{ij} = \left.\dfrac{\partial \mathbf G\left(\mathbf R_i(t) - \mathbf R_j(t)\right)}{\partial \mathbf R_i(t)}\right|_{t = 0},
\end{align}
and
\begin{align}\label{eq:T_ij}
&\mathbf T\left(\mathbf R_i(t) - \mathbf R_j(t)\right) = \nonumber \\
&\mathbf T\left(\mathbf R_i(0) - \mathbf R_j(0)\right) + \left.\dfrac{\partial \mathbf T\left(\mathbf R_i(t) - \mathbf R_j(t)\right)}{\partial \mathbf R_i(t)}\right|_{t = 0}  \cdot \left(\boldsymbol\xi_i(t) - \boldsymbol\xi_j(t)\right) \nonumber \\
& = \mathbf T_{ij} + \mathbf V_{ij} \cdot \left(\boldsymbol\xi_i(t) - \boldsymbol\xi_j(t)\right),\nonumber\\
&\text{ with } \mathbf T_{ij} = \mathbf T\left(\mathbf R_i(0) - \mathbf R_j(0)\right) \text{ and }\mathbf V_{ij} = \left.\dfrac{\partial \mathbf G\left(\mathbf R_i(t) - \mathbf R_j(t)\right)}{\partial \mathbf R_i(t)}\right|_{t = 0}.
\end{align}

Combining Eqs.~(\ref{eq:Cdot})-(\ref{eq:T_ij}) and using the fact that $\mathbf F_i^\text{d}(t)$ and $\mathbf F_i^\text{s}(t)$, the driving and the spring forces, sum to zero over the three bodies at each instant, we get
\begin{align}\label{eq:Cdotfull}
& \dot{\mathbf{C}}(t) = \dfrac{\left(1 - \sum\limits_{p = 1}^3\sum\limits_{m = 1}^\infty \bar{b}_p^m \sin(m\omega t + \phi_p^m)\right)}{\sum\limits_{j = 1}^3 a_j} \times \nonumber \\
& \sum\limits_{i = 1}^3\sum\limits_{j \neq i}^3 \left\{\left(a_i  + \sum\limits_{n = 1}^\infty b_j^n \sin(n\omega t + \phi_j^n)\right)\right .\cdot \left .\left(\mathbf T_{ij} + \left(\boldsymbol\xi_i(t) - \boldsymbol\xi_j(t)\right)\cdot\mathbf V_{ij}\right)\right.\cdot\nonumber\\
& \left .\ \ \ \ \ \ \ \ \ \ \ \ \left(\mathbf F_j^\text{d}(t) + \sum\limits_{k \neq j}^3 \left(\boldsymbol\xi_j(t) - \boldsymbol\xi_k(t)\right)\mathbf H_{jk}\right)\right\}.
\end{align}

Making use of the condition $b_i^n/|\boldsymbol\xi_j(t)| \ll 1$, we expand all terms in Eq.~(\ref{eq:v_def}) to the lowest surviving order in $\boldsymbol\xi_i(t)$ after inserting in it the expression for $\dot{\mathbf C}(t)$ from Eq.~(\ref{eq:Cdotfull}). Since the displacements $\boldsymbol\xi_i(t)$ directly arise from the driving forces $\mathbf F_j^\text{d}(t)$, this also means the lowest surviving order in $\mathbf F_j^\text{d}(t)$.~\cite{Felderhof:2006:PhysFluids} Moreover, since both the displacements and the driving forces are sinusoidal, it turns out that all terms in Eq.~(\ref{eq:v_def}) of the zeroth or first order in $\boldsymbol\xi_i(t)$ or $\mathbf F_j^\text{d}(t)$ average to zero over a cycle. Checking now for the second order terms, it can be easily shown that only the $n = 1$ terms for the surface fluctuation $\sum\limits_{n = 1}^\infty b_j^n  \sin(n\omega t + \phi_j^n)$ contribute, since all the other modes also evaluate to zero due to the orthogonality of the sinusoidal functions.

Therefore, given sufficiently weak periodic surface deformations of any functional form, only the driving frequency modes contribute to the swimming motion. The terms that survive in Eq.~(\ref{eq:Cdotfull}) are all sinusoidal functions, and the equation then becomes straightforward to solve, finally yielding an expression for $\mathbf v_\text{def}$ of the form
\begin{equation}\label{eq:v_def_final}
\mathbf v_\text{def} = \mathbf v_\text{rigid} + \sum\limits_{i = 1}^3\mathbf m_i b_i^1,
\end{equation}
where $\mathbf v_\text{rigid}$ is the velocity of a corresponding swimmer with all beads rigid,~\cite{Pande:2016:PNAS} and the coefficients $\mathbf m_i$ are independent of the shape deformation amplitudes $b_j^1$. The signs of the $\mathbf m_i$'s determine whether the swimming velocity is increased or decreased by the deformations of the beads, and we shall show by comparison to simulations that both the cases are possible. The full expression for $\mathbf v_\text{def}$ in Eq.~(\ref{eq:v_def_final}) is provided in the Electronic Supplementary Information (E.S.I.) in the form of a \emph{Mathematica} file. 

\subsection{Phase diagram}

\begin{figure}
\centering
\includegraphics[width=.48\textwidth]{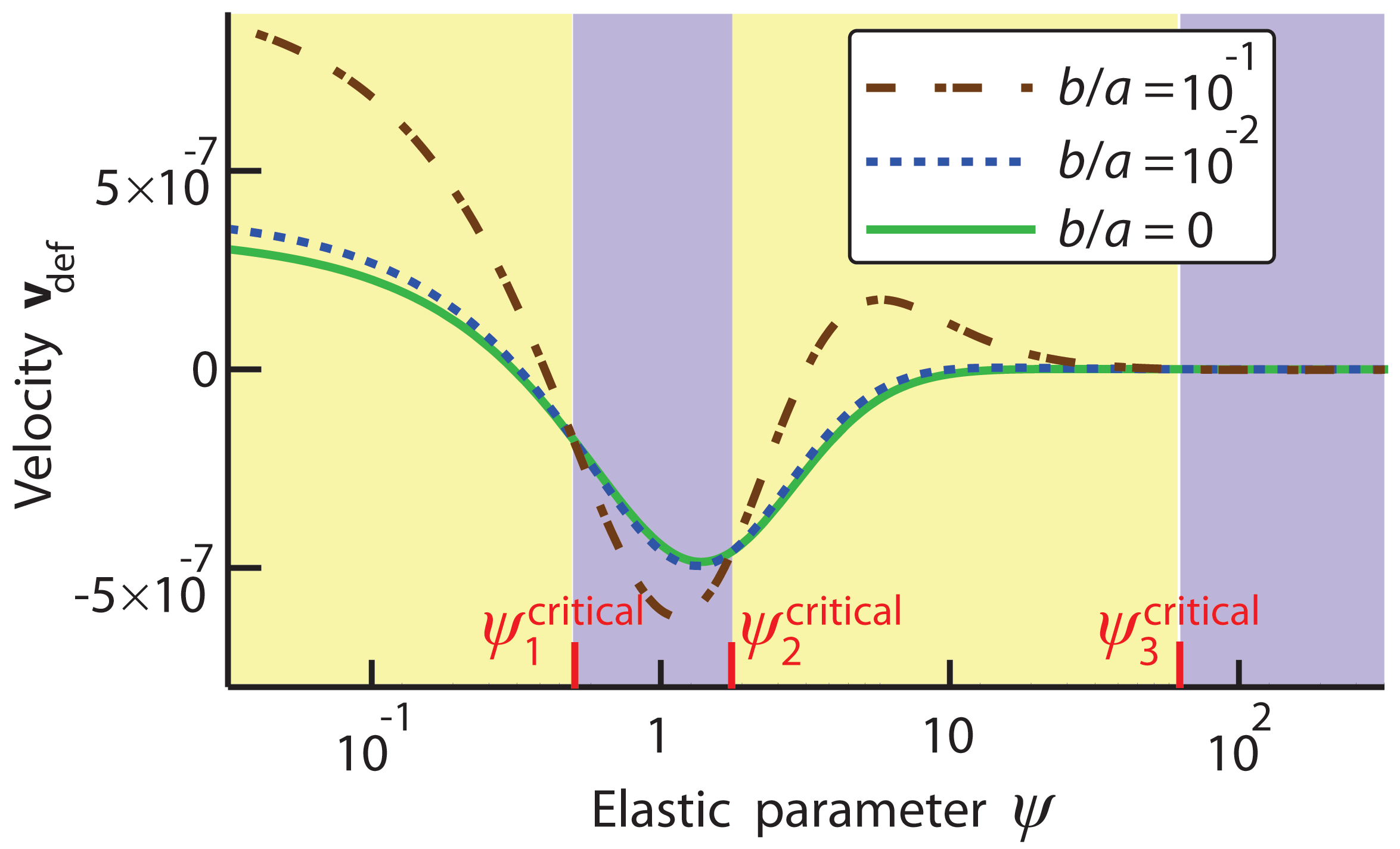}
\caption{(Colour online) Velocity vs. effective elastic parameter for swimmers with deformable beads. The light yellow and violet colours mark the deformability-enhanced and deformability-hindered regions, respectively. The solid green curve is for a swimmer with rigid beads.
}\label{fig:deformability_regimes}
\end{figure}

We now identify the two regimes of deformability-enhanced or deformability-hindered motion for the special case of identical beads deforming with equal oscillation amplitudes, \textit{i.e.}\ with $a_i = a$ and $b_i^1 = b$, for $i = 1,2,3$. While this is not a realistic situation since in actual swimming the different beads are expected to react differently to the fluid flow, such an analysis is instructive as one can readily identify the main factor that decides the deformability-based regime in which the motion occurs. (Moreover, a physical scenario wherein such an analysis becomes relevant is when the bead surfaces can be deformed due to external control but are resistant to the fluid flow itself. In this case, of course, the deformations can no longer be termed `passive'.)

In this special case, the sum of the three $\mathbf m_i$'s in Eq.~(\ref{eq:v_def_final}) can be redefined as $\mathbf m$, and 
we find that it is the spring constant $k$ which determines whether making the beads more deformable aids ($\mathbf m > \mathbf 0$) or opposes ($\mathbf m < \mathbf 0$) the swimming, independent of the oscillation amplitudes of the beads themselves. There are up to three critical values of $k$, which we denote as $k^{\text{critical}}_i$ ($i = 1, 2, 3$), that separate these deformability-enhanced and deformability-hindered regions. These regions are marked by violet and yellow colours, respectively, in 
Fig.~\ref{fig:deformability_regimes}, which shows a case in which all three $k^{\text{critical}}_i$ values are physically meaningful, \textit{i.e.}\ real and positive. 
For explicit expressions for $k^{\text{critical}}_i$, see the \emph{Mathematica} file in the E.S.I.


\begin{figure*}
\centering
\includegraphics[width=.8\textwidth]{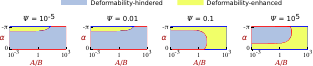}
\caption{(Colour online) Phase diagrams for swimmers with deformable beads, for different values of the forcing parameters $\alpha$, $A$ and $B$. 
}\label{fig:phase_diag}
\end{figure*}

In Fig.~\ref{fig:phase_diag} we present phase diagrams, again for $a_i = a$ and $b_i^1 = b$ (for $i = 1, 2, 3$), showing the relative extent of the deformability-enhanced and -hindered regimes for different driving force amplitude ratios $A/B$ and phase shifts $\alpha$, for a few values of the spring constant $k$. In each diagram, all the parameters apart from $A/B$ and $\alpha$ are kept constant. The diagrams show that for soft springs (small values of $k$), the swimmer becomes slower with increasing bead deformability for a majority of the phase space. As the springs become stiffer, the swimmer becomes more likely to benefit from an increase in the bead deformabilities. This happens because in the limit of perfectly stiff springs (\textit{i.e.}\ infinite $k$), a swimmer with rigid beads cannot swim as it is one contiguous, rigid body, and in this case deformable beads are necessary to gain the requisite degrees of mechanical freedom which may lead to motion.

\section{Lattice-Boltzmann simulations}
%
To check the results from our theory, we use the LB3D code \cite{Jansen:2011:PRE, Krueger:2013:EPJST} to run simulations of our deformable bead-spring swimmers for different parameter values. This code combines the lattice-Boltzmann method (LBM) for the fluid, the finite element method (FEM) for the deformations of the membranes, and the immersed boundary method (IBM) for the coupling between the fluid and the membranes. The lattice structure employed is the D3Q19 lattice and the collision operator is given by the Bhatnagar-Gross-Krook (BGK) model, both standard choices.

In all the simulations the driving forces (Eq.~(\ref{eq:dr_forces})) are applied on the centres of the beads, and the swimming stroke as well as the surface deformations of the beads emerge in response to the various forces acting on them (namely those due to the driving, the springs, the fluid, and, locally on a membrane element, due to other membrane elements). In all the simulations we stipulate the radii of the initially spherical beads to be $r = 5\Delta x$, where $\Delta x$ denotes one lattice cell length, and the mean centre-to-centre distance between two neighbouring beads to be $l = 36\Delta x$. The deformable bead membranes are modelled by meshes which have 720 triangular faces each, and are generated by successively subdividing an initially coarse icosahedron mesh. 

When the $i^\text{th}$ bead deforms, it gains an energy of deformation $W_i$ in the simulations, which is specified to be
\begin{equation}\label{eq:W_i}
W_i = W^\text{S}_i + W^\text{B}_i + W^\text{V}_i,
\end{equation}
where the superscripts S, B and V label the energy contributions due to strain, bending, and volume change, respectively. The strain energy is as given by the Skalak model,~\cite{Skalak:1973:BiophysJ} and has two associated stiffness moduli, the shear modulus $k^\text{S}_i$ and the area dilation modulus $k^\alpha_i$. 
The bending energy $W^\text{B}_i$ (with the associated bending modulus $k^\text{B}_i$) is significant in the presence of strong local curvatures, but turns out not to be important in our simulations due in part to the initially spherical shape of the beads. The global volume energy $W^\text{V}_i$ (with a corresponding volume modulus $k^\text{V}_i$) imposes restrictions on the change of bead volume, with this energy being the lowest if the volume is at its equilibrium value. Note that in our simulations the global surface area of the beads is not conserved. For more details on the different deformation energy parameters, see Kr\" uger \emph{et al}.~\cite{Krueger:2011:CMA} 


We elect to restrain the total volume of the beads in the simulations, setting $k^\text{V}_i = 1$, which results in volume deviations smaller than $0.01\%$. The bending modulus $k^\text{B}_i$ turns out not to affect the simulation results much, due to the initially spherical shape of the beads and the fact that they do not undergo very strong deformations. Specifically, we find that changing $k^\text{B}_i$ by 5 orders of magnitude changes the swimming velocity by less than $1\%$ (data not shown). Accordingly, in all the simulations discussed in this paper, we keep the value of the bending modulus fixed at $k^\text{B}_i = 10^{-3}$. In the different simulation sets, therefore, only the shear moduli $k^\text{S}_i$ and the area moduli $k^\alpha_i$ are changed, but always under the condition that $k^\text{S}_i = k^\alpha_i \ \forall\ i$. We stick to this condition in order to reduce the parameter space--physically it means that the membranes respond equally strongly to both deformation and dilation. For different beads in a swimmer, these two stiffness moduli are allowed to be different ($k^\text{S}_i \neq k^\text{S}_j$). In all the results discussed in this paper, we refer only to a variation in $k^\text{S}_i$ in the simulations, with the variation in $k^\alpha_i$ being implied.

The beads contain fluid which has the same density and viscosity as the external fluid. In the simulations, the kinematic viscosity $\nu$ of the fluid (related to the dynamic viscosity $\mu$ as $\nu = \mu/\rho$, where $\rho$ is the fluid density) is kept at $0.2$ in lattice units. Since $\nu = (2 \tau - 1) / 6$, where $\tau$ is the lattice-Boltzmann relaxation parameter, this corresponds to $\tau = 1.1$, which in previous work we have found to be the optimal value for highly accurate swimmer simulations compared to theory when the beads of the swimmer are rigid.~\cite{Pande:2016:PNAS}

\begin{figure}
\centering
\includegraphics[width=.48\textwidth]{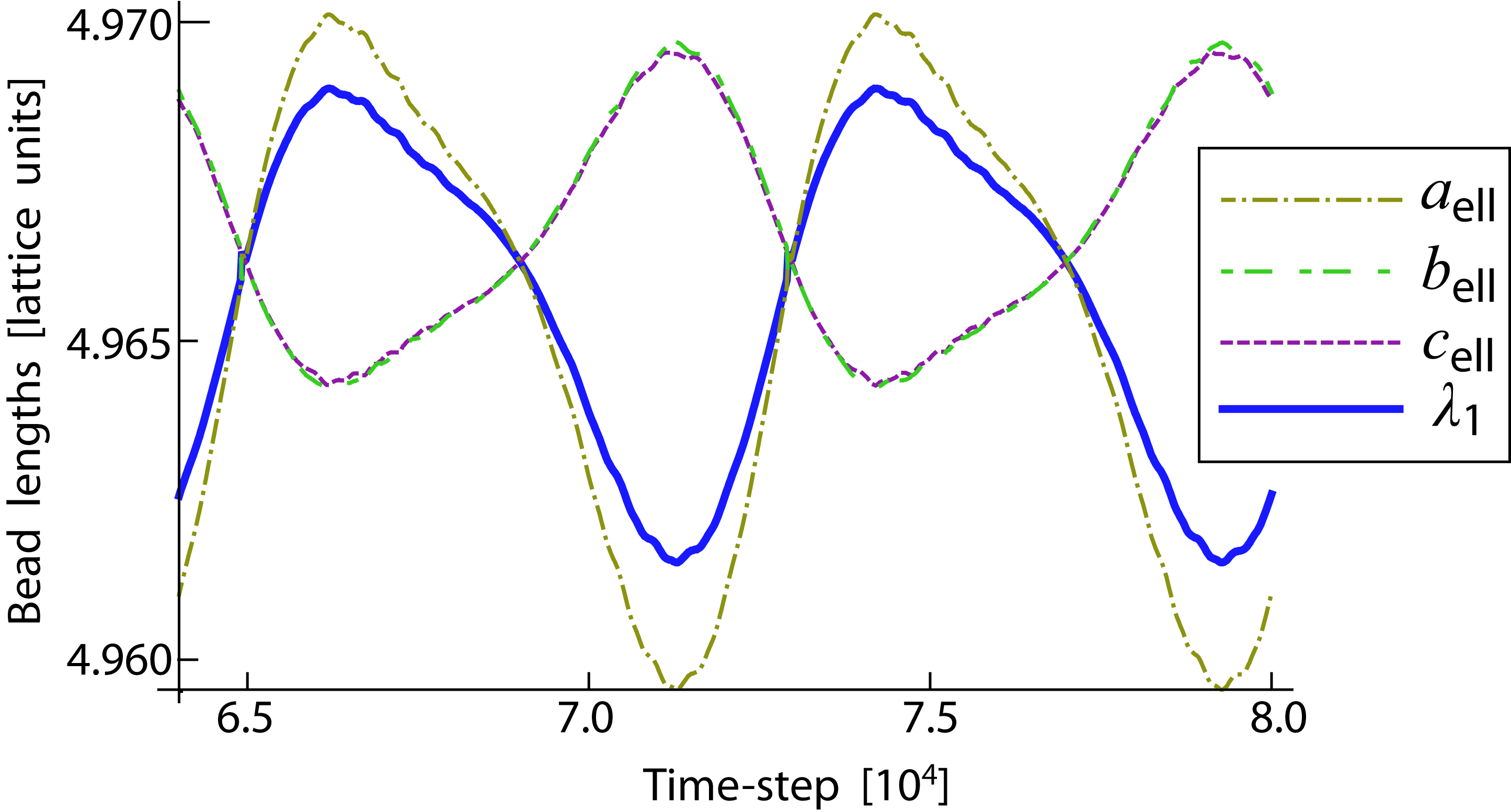}
\caption{(Colour online) Different semi-axes of the bounding ellipsoid of bead 1, and the resultant effective friction coefficient $\lambda_1$, for a simulation with $k_\text{s} = 0.1$, $A/B = 20$ and $\alpha = \pi/2$.}\label{fig:semi-axes}
\end{figure}


To be able to compare the results of the simulations with those that our theory gives, an input we need for the latter is the time-dependent effective friction coefficient $\lambda_i(t)$ of each deformable bead (Eq.~(\ref{eq:lambda_i})). For weak bead deformability, which our theory assumes, the initially spherical beads retain a convex shape throughout and can be approximated as ellipsoids. We can therefore fit the instantaneous shape of the $i^\text{th}$ bead ($i = 1, 2, 3$) in a simulation to a bounding ellipsoid and determine its three semi-axes $s_i^1$, $s_i^2$ and $s_i^3$ (Fig.~\ref{fig:semi-axes}). We do this by comparing the bead shape to an ellipsoid of the same inertia tensor, a procedure described in detail in Kr\" uger \emph{et al.}~\cite{Krueger:2011:CMA} 
Due to the axisymmetry of the problem, at each moment two of the semi-axes in a bead are equal, so that the bounding ellipsoids are always spheroids (\textit{i.e.}\ ellipsoids of revolution), although these spheroids can change from prolate to oblate and back within a swimming cycle. The effective friction coefficients of the beads at each instant are then found using Perrin's formulas,~\cite{Perrin:1934:JPhysRadium, Pande:2015:SoftM} and fitting with Eq.~(\ref{eq:lambda_i}) provides the different parameters of deformation. As an example, Fig.~\ref{fig:semi-axes} shows the three semi-axes and the resulting effective friction coefficient of bead 1 in one of our simulations (presented in Fig.~\ref{fig:vel_comparison} (a)).

\subsection{Comparison with theory}

\begin{figure}
\centering
\includegraphics[width=.4\textwidth]{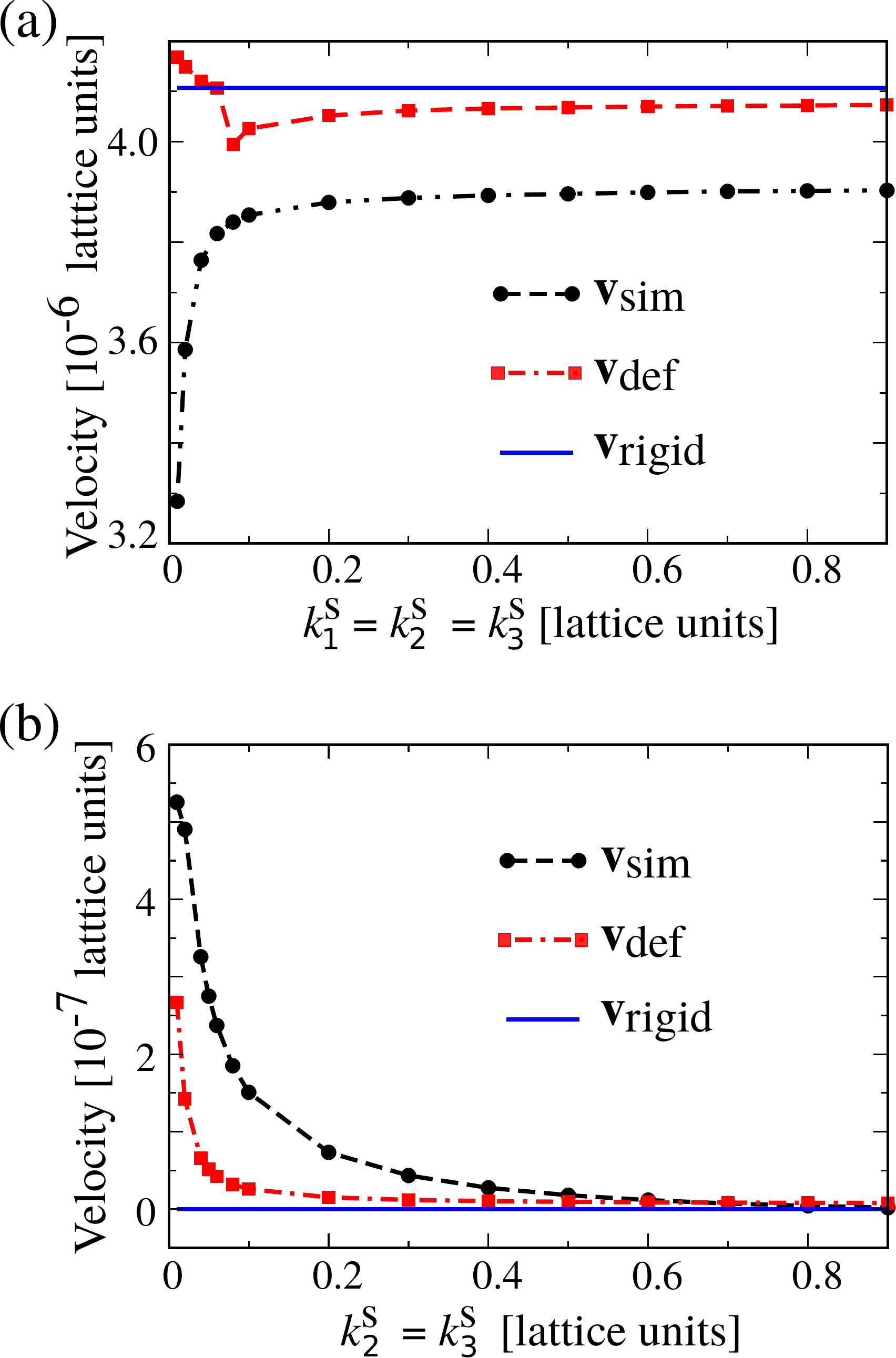}
\caption{(Colour online) Comparison of swimmer velocity found from simulations and theoretical velocity expressions assuming rigid and deformable beads, for changing shear moduli $k^\text{S}_i$ of the beads. In (a), the $k^\text{S}_i$ values of all the beads are changed together, and the force ratio and the force phase difference respectively are $A/B = 20$ and $\alpha = \pi/2$. In (b), $k^\text{S}_1$ is held constant at 1 while the $k^\text{S}_i$ values for $i = 2, 3$ are changed together, and the force ratio and the force phase difference are respectively $A/B = 1$ and $\alpha = 0$. Note that the forcing is symmetric in (b), meaning that in the case of rigid beads ($k^\text{S}_2 = k^\text{S}_3 \gtrsim 0.8$) there is no swimming.}\label{fig:vel_comparison}
\end{figure}

%
As anticipated by the theory, the simulations provide evidence of both the deformability-hindered and -enhanced regimes existing for the swimming velocity. We present instances of both these cases in Fig.~\ref{fig:vel_comparison}, where the swimming velocity is shown as a function of the bead shear moduli $k^\text{S}_i$ for two different sets of driving force amplitude ratios $A/B$. 
The black curves in the plots show the simulation results ($\mathbf v_\text{sim}$) and the red curves the velocities found using our theory ($\mathbf v_\text{def}$). In addition, the blue solid curves show the theoretically-expected velocities for similar swimmers but with rigid beads ($\mathbf v_\text{rigid}$),~\cite{Pande:2015:SoftM} and are naturally constant-valued. 

In part (a) of Fig.~\ref{fig:vel_comparison}, the force amplitudes $A$ and $B$ are related as $A = 20 B$, and the phase difference between the two is $\alpha = \pi/2$. In this set, the shear moduli of all the beads in the swimmer are changed together and are always equal ($k^\text{S}_i = k^\text{S}_j$ $\forall$ $i,j$). It is seen that in the simulations the swimming velocity $\mathbf v_\text{sim}$ goes down when $k^\text{S}_i$ decreases. Significantly, the theoretical $\mathbf v_\text{def}$ curve echoes this decrease in the swimming velocity for decreasing $k^\text{S}_i$, for $k^\text{S}_i > 0.08$. For smaller shear moduli (which correspond to larger bead deformabilities), the theoretical analysis breaks down as it is based on the weak-deformability limit (which is why the black and the red curves diverge from each other for $k^\text{S}_i < 0.08$). 
We also find that for $k^\text{S}_i > 0.08$, the theory for deformable beads performs better in comparison to the simulations than the rigid-bead theory ($\mathbf v_\text{rigid}$).

In part (b) of Fig.~\ref{fig:vel_comparison}, the force amplitudes $A$ and $B$ are kept equal, and the phase difference between the two is $0$. This is thus a case of symmetric driving, and for rigid beads this results in no net swimming. To obtain swimming behaviour, we let one bead in the swimmer remain rigid ($k^\text{S}_1 = 1.0$), while the shear moduli of the other two beads are varied together ($k^\text{S}_2 = k^\text{S}_3$), thus breaking the symmetry in the setup and resulting in swimming. Since the velocity is $0$ for rigid beads ($k^\text{S}_i = 1.0\ \forall\ i$), therefore for at least some range of $k^\text{S}_i \le 1$ we must obtain the deformability-enhanced regime, 
and this is precisely what the simulations show ($\mathbf v_\text{sim}$) as well as what the deformable bead theory ($\mathbf v_\text{def}$) gives. 
In this case we find that the theoretical $\mathbf v_\text{def}$ curve shows relatively good agreement with the simulations ($\mathbf v_\text{sim}$ curve) for $k^\text{S}_2 = k^\text{S}_3 > 0.01$.

It is pertinent to reiterate here the importance of the weak-deformability limit in our theory. Working within this limit allows us to treat the problem in a way similar to that for rigid beads with the deformation of the beads only affecting their effective friction coefficients $\lambda_i(t)$, without having an impact (or only an indirect one) on the fluid flow. This extracts as its price the restriction of our analytical approach to relatively large $k^\text{S}_i$ values, when the beads are close to rigid. This is also the reason why the $\mathbf v_\text{def}$ curves are a small correction to the $\mathbf v_\text{rigid}$ curves in both the parts of Fig.~(\ref{fig:vel_comparison}), without providing the precise velocity values as found in the simulations. Qualitatively, however, in terms of identifying the rise or fall of the velocity with the deformability, the theory reproduces the simulation trends faithfully for all the parameter sets investigated. 

\section{Conclusion}

In this work we have studied how the deformability of a mechanical microswimmer's body influences its motion. In particular we have considered the question of whether having a soft body, yielding to the surrounding fluid flow by undergoing shape changes, is beneficial to the motion of a swimmer for which the main mechanism for motility is independent of these shape changes. As our model we have picked a bead-spring swimmer, where the motion mechanism is the contraction and expansion of the springs due to driving forces, and where the beads are deformable such that they alter their shapes continuously as the swimmer makes its way through the fluid. We have found an analytical expression for the swimming velocity starting from a knowledge of the driving forces and the deformations of the beads. This analysis, predicated on the assumption of small body deformations such that the fluid flow is not affected by the fluctuating bead surfaces, predicts that the swimming velocity may both rise and fall with an increase in the body deformability. An important factor in determining which of the two kinds of behaviour is seen appears to be the elasticity characterising the dominant swimming mechanism, which for our swimmers translates to the spring constant.

Using lattice-Boltzmann immersed boundary simulations of the swimmer, we have confirmed this picture of the velocity being affected in opposite ways by the body deformability, depending on the different parameters. In all the cases that we have checked, our theoretical model reproduces correctly the velocity trends with increasing deformability and improves considerably upon the velocity values predicted by a theoretical model which only considers rigid beads in the swimmer.

Our study 
provides a possible explanation for why some microorganisms like euglenids undergo body shape changes while swimming, even though they do not apparently need to for their motion is driven by flagellar beating. We suggest that in the appropriate parameter ranges, such shape changes could enhance the motility. Our results are also useful for artificial microswimmers, since they show that the efficiency of these devices may be increased by exploiting body deformability appropriately. 
It is hoped that a third utility of our work will be in its analytical treatment of what is in general a very difficult problem, namely the deformations and consequent effect on swimming of a non-rigid membrane through a fluid, and that in the future this can be extended to a model where these deformations are calculated \textit{a priori}, at least in the small deformation limit.

\emph{Acknowledgment.}---A.-S.~S.~and J.~P.~are grateful to KONWIHR ParSwarm and ERC-2013-StG 337283 MembranesAct grants for financial support. 
 J.~H.~was supported by NWO/STW (Vidi grant 10787) and L.~M.~by the DAAD through a RISE scholarship. T.~K.~acknowledges support by the University of Edinburgh through the award of a Chancellor's Fellowship.

\bibliography{rsc3_for_arXiv} 
\bibliographystyle{rsc3}

\end{document}